\documentclass[twocolumn]{aastex62}
\usepackage{array} 
\usepackage{amsmath}
\usepackage{wasysym}

\newenvironment{tightcenter}{%
	\setlength\topsep{0pt}
	\setlength\parskip{0pt}
	\begin{center}
}{%
 	\end{center}
}

\newcommand\chandra{{\it Chandra}}

\newcommand\ciao{CIAO}
\newcommand\sherpa{Sherpa}

\newcommand\caldb{CALDB}

\newcommand\hst{{\it HST}}

\shorttitle{Superluminal Motion in X-Ray Jet of M87}

\shortauthors{Snios et al.}

\begin{document}

\title{Detection of Superluminal Motion in the X-Ray Jet of M87}

\author{Bradford Snios} 
\affil{Harvard-Smithsonian Center for Astrophysics, 60 Garden St, Cambridge, MA 02138, USA}
\author{Paul E. J. Nulsen} 
\affil{Harvard-Smithsonian Center for Astrophysics, 60 Garden St, Cambridge, MA 02138, USA}
\affil{ICRAR, University of Western Australia, 35 Stirling Hwy, Crawley, WA 6009, Australia}
\author{Ralph P. Kraft}
\affil{Harvard-Smithsonian Center for Astrophysics, 60 Garden St, Cambridge, MA 02138, USA}
\author{C. C. Cheung}
\affil{Space Science Division, Naval Research Laboratory, Washington, DC 20375, USA}
\author{Eileen T.  Meyer}
\affil{Department of Physics, University of Maryland Baltimore County, Baltimore, MD 21250, USA}
\author{William R. Forman}
\affil{Harvard-Smithsonian Center for Astrophysics, 60 Garden St, Cambridge, MA 02138, USA}
\author{Christine Jones}
\affil{Harvard-Smithsonian Center for Astrophysics, 60 Garden St, Cambridge, MA 02138, USA}
\author{Stephen S. Murray}
\altaffiliation{Steve Murray passed away on 2015 August 10. 
	His \\ contributions to X-ray astronomy will always be remembered.}
\affil{Harvard-Smithsonian Center for Astrophysics, 60 Garden St, Cambridge, MA 02138, USA}

\begin{abstract}
\chandra{} HRC observations are investigated for evidence of proper motion and brightness changes in the \mbox{X-ray} jet of the nearby radio galaxy M87. Using images spanning 5 yr, proper motion is measured in the \mbox{X-ray} knot HST-1, with a superluminal apparent speed of $6.3 \pm 0.4 c$, or $24.1 \pm 1.6\rm\ mas\ yr^{-1}$, and in Knot D, with a speed of $2.4\pm 0.6c$. Upper limits are placed on the speeds of the remaining jet features. The X-ray knot speeds are in excellent agreement with existing measurements in the radio, optical, and ultraviolet. Comparing the X-ray results with images from the {\it Hubble Space Telescope} indicates that the X-ray and optical/UV emitting regions co-move. The X-ray knots also vary by up to 73\% in brightness, whereas there is no evidence of brightness changes in the optical/UV.  Using the synchrotron cooling models, we determine lower limits on magnetic field strengths of  $\sim$\,$420~\mu \rm G$ and $\sim$\,$230~\mu \rm G$ for HST-1 and Knot A, respectively, consistent with estimates of the equipartition fields.  Together, these results lend strong support to the synchrotron cooling model for Knot HST-1, which requires that its superluminal motion reflects the speed of the relativistic bulk flow in the jet.
\end{abstract}


\section{Introduction}
\label{sect:intro}

Jetted outflows from central active galactic nuclei (AGNs) are a well-established phenomenon in extragalactic radio sources \citep[e.g.,][]{Blandford1974, Hargrave1974, Scheuer1982, Begelman1984}. Jets are known to propagate with bulk relativistic motion over kiloparsec-scale distances \citep{Begelman1984, Bridle1984}, and an increasing number of jets have been detected in wavelength bands other than radio, including the X-ray \citep[and references therein]{Marshall2018}. Direct observation of apparent jet velocities via proper motion studies may be used to place valuable constraints on line-of-sight angles and jet flow kinematics \citep{Vermeulen1994, Lister2013}. 

To date, the most comprehensive data on jet motions come from studies of the nearest sources, particularly the Fanaroff--Riley type-I \citep[FR\,I;][]{Fanaroff1974} radio galaxies M87 \citep{Reid1989, Biretta1995, Cheung2007, Kovalev2007} and Centaurus\,A \citep{Goodger2010, Muller2014}. M87 is a nearby source, at a distance of only 16.7~Mpc  \citep{Blakeslee2009}, that hosts a bright jet spanning 20\arcsec, $\sim$1.6~kpc projected, from the central AGN of the system. Kiloparsec-scale, knotted structure has been directly observed within the jet using radio, optical, and X-ray observations \citep[e.g.,][]{Owen1989, Sparks1996, Harris1997, Perlman2001, Marshall2002, Perlman2005, Meyer2013}.

Previous measurements of proper motions in M87 have detected sub-relativistic speeds within a parsec of the AGN using the Very Long Baseline Interferometry \citep[VLBI;][]{Reid1989, Kovalev2007, Mertens2016, Walker2018}. Superluminal motions on larger-scales were detected from the VLBI \citep{Cheung2007, Giroletti2012}, the Very Large Array \citep[VLA;][]{Biretta1995}, and the {\it Hubble Space Telescope} \citep[\hst;][] {Biretta1999, Meyer2013}. While proper motions have yet to be detected at higher energies, measuring the kinematics of X-ray-emitting regions potentially provides the best probe for sites of particle acceleration becauser high energy radiation is produced by more energetic particles. Recent work from \cite{Snios2019} measured X-ray proper motions for the first time in the jet of Centaurus\,A, at a distance of 3.8 Mpc \citep{Harris2010}, by applying new methods to analyze \chandra{} observations. We may apply those techniques to measure proper motions in the X-ray jet of M87.

In this work, we investigate proper motions and brightness variations within the X-ray jet of M87 using two \chandra{} observations separated by 5 yr. The remainder of the paper is structured as follows. Section~\ref{sect:observation} discusses the details of the \chandra{} observations and the corresponding data reduction. In Section~\ref{sect:diffmap}, difference maps are generated from the coaligned observations and statistical uncertainties are quantified. Evidence of proper motion for the innermost knots of the X-ray jet are presented in Section~\ref{sect:motion}. In Section~\ref{sect:discuss}, the proper motion results are compared with measurements from other wavelengths. Lastly, brightness changes of the knots are utilized to investigate both the primary mechanism for fading and the magnetic field strengths of the knots.

\section{Data Acquisition and Reduction} 
\label{sect:observation}

The jet of M87 is known to be a bright X-ray source. In existing \chandra{} observations with the Advanced CCD Imaging Spectrometer (ACIS), several features of the jet, including the AGN and HST-1, are significantly affected by pile-up ($>$ 0.25 counts frame$^{-1}$; \citealt{Marshall2002,Wilson2002, Harris2003}). ACIS observations were initially considered for our analysis, but we could not achieve the required astrometric alignment accuracy between epochs due to the pile-up issues and a lack of background point sources against which to align. To avoid these issues, we instead focused our analysis on observations taken with the \chandra{} High Resolution Camera (HRC-I). HRC-I is a microchannel plate (MCP) detector that possesses the highest spatial resolution for \chandra{}, at a pixel size of 0.132\arcsec{} and full width at half maximum (FWHM) of 0.4\arcsec. HRC-I is less susceptible to pile-up due to its detector design. The large field of view of HRC-I also ensures numerous X-ray point sources surrounding M87 are detected in each observation \citep{Jordan2004}, where the background sources may collectively be used for astrometric alignments between different epochs. HRC-I is therefore an ideal instrument for X-ray proper motion studies.

\begin{figure*}
	\begin{tightcenter}
	\includegraphics[width=0.50\textwidth]{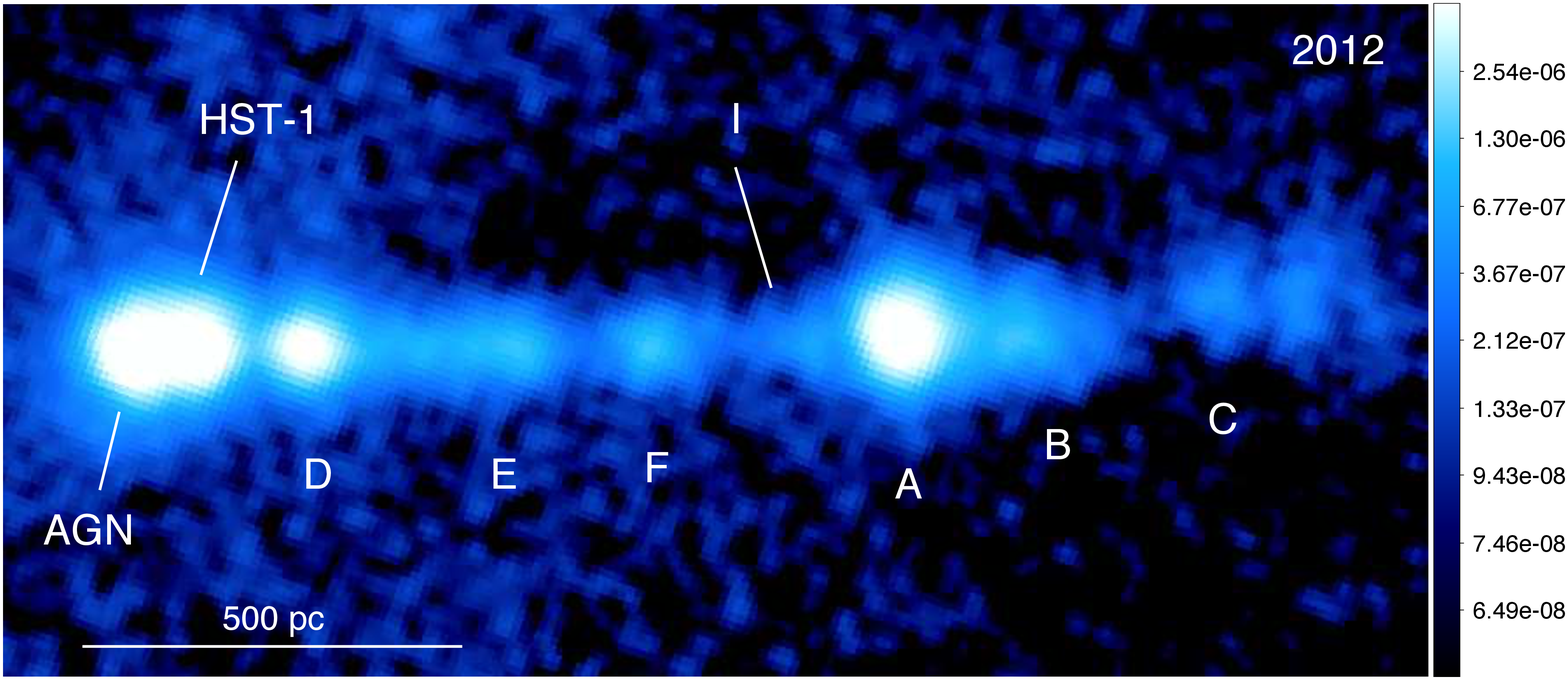}\includegraphics[width=0.50\textwidth]{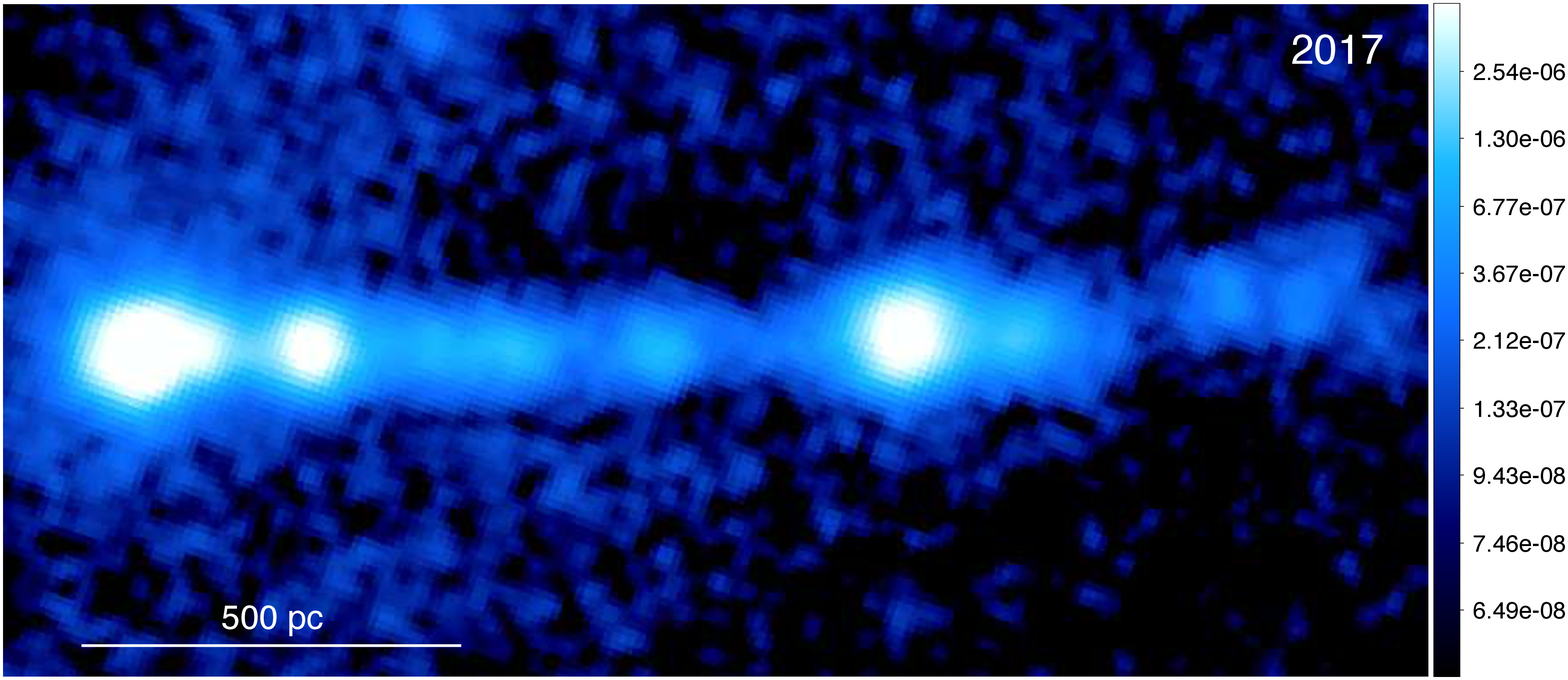} \\
	\includegraphics[width=0.50\textwidth]{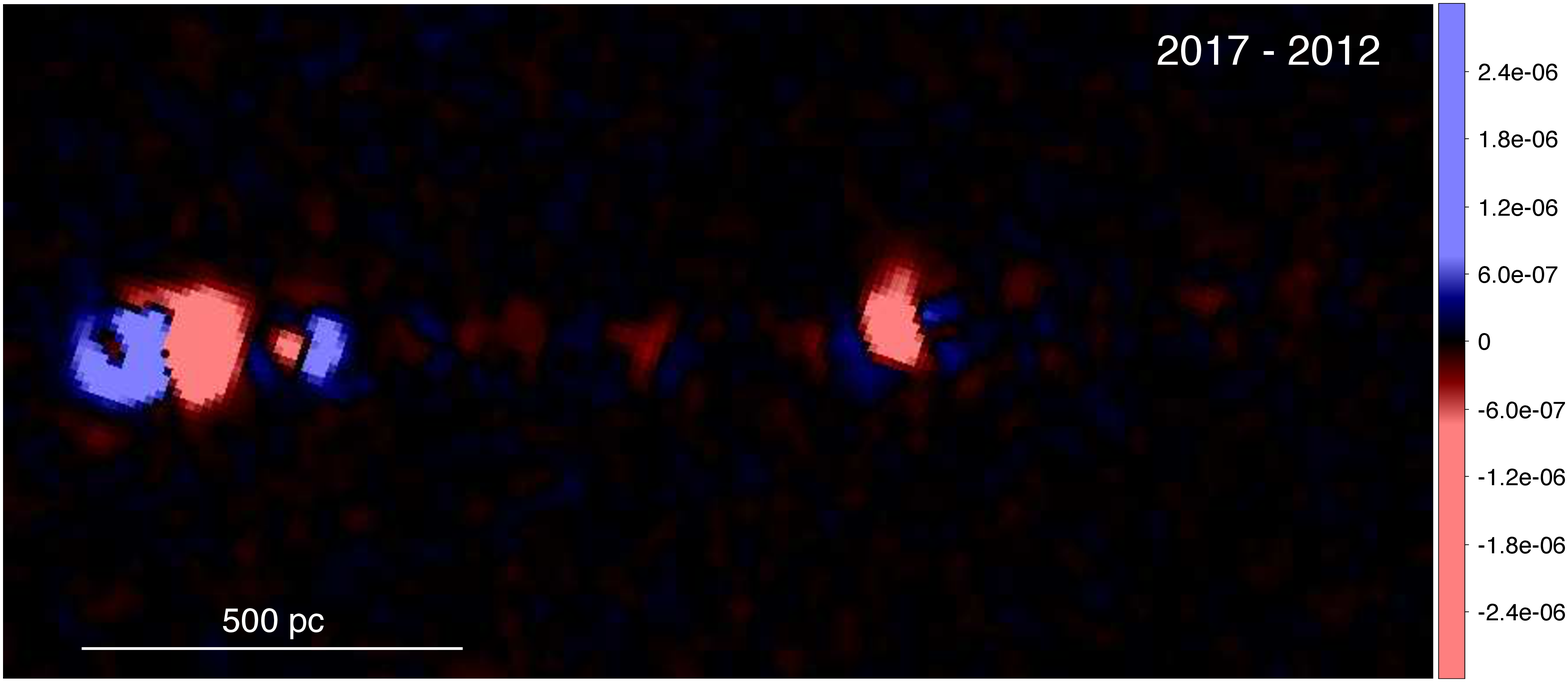}\includegraphics[width=0.50\textwidth]{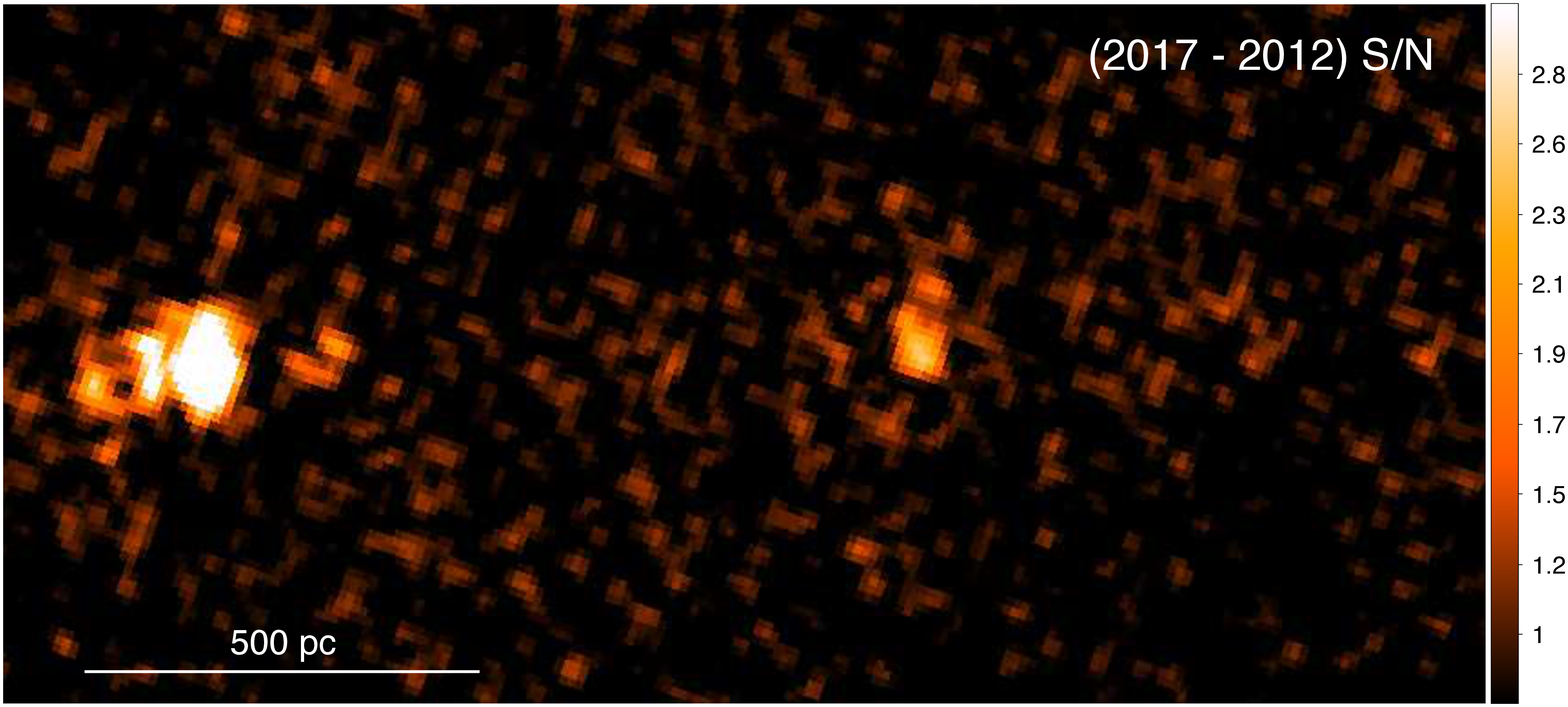} 
	\end{tightcenter} 		
      \caption{\chandra{} HRC-I 0.08--10.0 keV images (upper panels), difference map 
     		(lower left panel), and S/N of the difference map (lower right panel) of the 
		M87 jet. The images are binned on a scale of $0.132\arcsec \rm\ pix^{-1}$ 
		and are smoothed with a 2 pixel RMS Gaussian. For the difference map, 
		the red regions correspond to brighter areas in the 2012
		dataset, while the blue regions correspond to brighter areas 
		in the 2017 dataset. HRC-I and difference map images are in units of 
		photons cm$^{-1}$ s$^{-1}$.}
	\label{fig:m87}
\end{figure*}

M87 was initially observed with \chandra{} HRC-I in 2012 April with the aimpoint centered on Knot F of the jet (see Figure~\ref{fig:m87}) to maximize the spatial resolution of the system. A follow-up HRC-I observation using the same telescope configuration and aimpoint was performed in 2017 March with the explicit purpose of probing for motion and variations of brightness in the system over the 5 yr time span. Both observations were reprocessed using \ciao{} 4.10 with \caldb{} 4.7.8 \citep{Fruscione2006}. Background flares were removed using the \ciao{} routine {\tt deflare}. The resulting cleaned exposure times and further details on the selected observations are shown in Table~\ref{table:obs}. The count rate for the 2017 dataset was shown to be similar to the 2012 count rate after accounting for changes in the \chandra{} response over this interval, ensuring similar statistics for both observations. Exposure-corrected flux images were generated for each observation with {\tt fluximage}, and the latest HRC-I effective areas from \caldb{} were incorporated to correct for differences in effective area, quantum efficiency, and UV Ion-shield filter transmission between epochs. The final images are in units of $\rm photon\ cm^{-2}\ s^{-1}$. 

\begin{table}
	\caption{\textit{Chandra} Observations of M87 Used}
	\label{table:obs}
	\begin{tightcenter}
		\begin{tabular}{ c c c c }
		\hline
		\hline
		ObsID & Instrument & Date & $t_{\rm exp}^{a}$ (ks)\\
		\hline
			13515 & HRC-I & 2012 Apr 14 & 74.3 \\
			18612 & HRC-I & 2017 Mar 2 & 72.5 \\
		\hline
	\end{tabular}
	\end{tightcenter}
	{${}^{a}$Net exposure after background flare removal}
\end{table}

Using the method of our previous proper motion study \citep{Snios2019}, the cross-correlation between exposure-corrected images was used to determine and correct the small astrometric offset between the data sets. A rectangular region of $450\arcsec \times 375\arcsec$ centered on the AGN of M87 was extracted for each epoch. The jet was masked out to prevent jet features from having any impact on the resulting image alignment. Based on previous measurements of the AGN's motion \citep{Kovalev2007}, its shift would be unresolved in this study and was therefore left unmasked. In total, 25 unique point sources present in both epochs, including the AGN, were used to align the two observations. The two-dimensional cross-correlation function of the images was fitted with a two-dimensional Lorentzian profile to determine their relative offset. The offset was corrected using the \ciao{} task {\tt wcs\_update}. 

To test the systematic error in the image alignments, a comparison of the relative AGN centroid position was performed using the \ciao{} {\tt dmstat} routine. The AGN position was found to agree within $0.005\arcsec$ between the epochs. For comparison, proper motions in M87 at a speed of $c$ would produce a total shift of $0.019\arcsec$ over the 4.92 yr time span. Since superluminal proper motion in M87 greater than 6$c$ has previously been observed from optical and radio observations \citep{Biretta1999, Cheung2007, Meyer2013}, both images were concluded to be coaligned precisely enough for the analysis. The exposure-corrected images of M87, with labels for all features in the jet, are shown in Figure~\ref{fig:m87}. In this work, we define a knot as any feature in the jet that is distinguished by a factor of at least 2 in surface brightness from its surroundings and has a radius $< 2\arcsec$. We define the jet axis direction at a positional angle projected on the sky of 289$^{\circ}$. 

\section{Difference Maps}
\label{sect:diffmap}

Using reprocessed and coaligned images from Section~\ref{sect:observation}, a difference map of the two epochs was generated with the {\tt dmimgcalc} \ciao{} routine. The final images and difference map are shown in Figure~\ref{fig:m87}. Notable changes in brightness and morphology are observed in the AGN, HST-1, Knot D, and Knot A. Fainter knots in the jet, such as Knots E and F, also vary, though at a reduced statistical significance.  

To calculate the statistical and systematic uncertainties present in the difference map, we followed the procedure outlined in \cite{Snios2019}. The signal-to-noise ratio (S/N) for the difference map was measured as 
\begin{equation} 
{\rm S/N} = \frac{|c_2 N_2 - c_1 N_1|}{\sqrt{c_1^2 N_1 + c_2^2 N_2}}\,,
\label{eq:1} 
\end{equation} 
where the raw counts for a pixel in the two epochs are $N_1$ and $N_2$, and the corresponding exposure corrections are $c_1$ and $c_2$. The computed S/N map is shown in Figure~\ref{fig:m87}. In addition, the integrated S/N was determined for each knot in the jet. A region surrounding each knot was defined, and the integrated squared residual was defined as 
\begin{equation} 
R = \sum\limits_{i}^{N_{P}} ({\rm S/N})_{i}^2\,, 
\end{equation} 
where $({\rm S/N})_i$ is the S/N of Equation~\ref{eq:1} for the $i^{\rm th}$ pixel and $N_P$ is the total number of pixels within the region. See Table~\ref{table:diffbright} for values of $R$ and $N_P$ for each region. Values of $R$ should have a $\chi^2$ distribution with $N_P$ degrees of freedom, allowing us to determine the significance of changes in any region.  The values of the integrated S/N, $R$, for the AGN, HST-1, Knot D, and Knot A are all well in excess of the $3\sigma$ threshold for a $\chi^2$ distribution.  For the remaining jet features, the values of the integrated square residual, $R$, fall below the $3\sigma$ limit, consistent with their lack of prominence in the S/N map and difference map.

\begin{table*}
	\caption{Changes in X-Ray Brightness of Jet Features}
	\label{table:diffbright}
	\begin{tightcenter}
		\begin{tabular}{ c c c c c }
		\hline
		\hline
		Feature & $R/N_{P}$ & \multicolumn{2}{c}{Flux$^{a}$ 
			(10$^{-5}$ photons cm$^{-2}$ s$^{-1}$)} & Change in \\
		\cline{3-4}
		& [$({\rm S/N})^2/\rm pixels$] & 2012 Epoch & 2017 Epoch 
			& Flux$^{b,c,d}$ \\
		\hline
		AGN & 664.0/151 & $86.5 \pm 0.8$ & $106.3 \pm 0.9$ 
			&  $+23 \pm 5\%$ \\
		HST-1 & 2838.7/152 & $73.4 \pm 0.7$ & $20.4 \pm 0.4$ 
			&  $-73 \pm 5\%$ \\ 
		D & 220.0/134 & $31.5 \pm 0.5$ & $34.5 \pm 0.5$ 
			&  $+10 \pm 6\%$ \\
		E & 136.9/146 & $9.6 \pm 0.3$ & $9.4 \pm 0.3$ 
			& $-2 \pm 8\%$ \\
		F & 151.8/112 & $6.4 \pm 0.2$ & $5.8 \pm 0.2$ 
			& $-10 \pm 9\%$ \\ 
		I & 42.5/49 & $1.4 \pm 0.1$ & $1.2 \pm 0.1$ 
			& $-12 \pm 20\%$ \\
		A & 459.6/259 & $65.7 \pm 0.4$ & $57.7 \pm 0.7$ 
			& $-12 \pm 5\%$ \\ 			
		B & 209.3/220 & $11.3 \pm 0.3$ & $11.7 \pm 0.3$ 
			& $+4\pm 8\%$ \\
		C & 310.8/296 & $7.1 \pm 0.2$ & $6.3 \pm 0.2$  
			& $-13 \pm 11\%$ \\		
		\hline
	\end{tabular}
	\end{tightcenter}
	{${}^{a}$Flux emitted over the 0.08-10.0 keV energy band \\
	${}^{b}$Defined as ($F_{2017}$ - $F_{2012}$)/ $F_{2012}$, where is $F$ is flux \\
	${}^{c}$A positive value signifies an increase in the 2017 epoch; a negative value signifies a decrease \\
	${}^{d}$Uncertainties include statistical and systematic errors, as discussed in Section~\ref{sect:diffmap}}
\end{table*}

A significant squared residual, $R$, could reflect a change in the brightness of a jet feature, a change in its position, or both. Here we consider changes in brightness. From \chandra{} calibrations, the HRC-I instrument is known to possess a systematic count rate error $<$\,5\%\footnote{See Section 7.9 of the ``Proposers' Observatory Guide" \\ \url{http://cxc.harvard.edu/proposer/POG/html/chap7.html}}. To verify which observed variations exceeded the uncertainty threshold, the percentage change in brightness was determined for each feature in the X-ray jet. Regions were defined by surrounding each bright feature based on the knot definition criteria outlined in Section~\ref{sect:observation}, where regions were allowed to vary between epochs. Brightness was measured for each epoch over the \mbox{0.08--10.0} keV energy band of the HRC instrument. The diffuse jet emission was measured and subsequently subtracted from each result. Uncertainties were estimated based on Poisson noise and the systematic error of HRC. The brightness and brightness change for each examined feature are shown in Table~\ref{table:diffbright}. 

We conclude that the AGN has increased in overall brightness between the epochs to a $5\sigma$ significance, while HST-1 has decreased in brightness to a $14\sigma$ significance. Knots A and D were also found to decrease and increase in brightness over the examined time span, respectively, albeit to a marginal significance of $\sim$\,$2\sigma$. Variations in the brightness of the remaining knots were not significant individually, although the scatter in the brightness changes is consistent with more modest variations in these knots, too.

\section{Proper Motion}
\label{sect:motion}

\chandra{} HRC-I observations were examined for evidence of proper motion in the jet of M87. Any changes are expected to be subtle at the resolution of \chandra{} as movement at the speed of light over a 5 yr time span in M87 would produce a shift of only $\sim\,0.02\arcsec$. In the difference map (Figure~\ref{fig:m87}, bottom left), the outward shift of a knot with constant brightness would produce positive residuals (blue) at its outer margin and negative residuals (red) at its inner margin. These features can be seen in the difference image of Knot D, providing clear evidence that it has moved outward along the axis of the jet. No other knot shows such clear evidence of movement, although this may be attributed in part to significant changes in brightness, such as those seen in HST-1 and  Knot A. Other knots are too faint for the expected changes to produce statistically significant features in the difference image. In the following sections, we seek to quantify the motion of the knots through the use of more rigorous methods. 

\subsection{Motion Measurements from Cross-Correlation and Centroid Analyses}
\label{sect:motioncc}

\begin{table*}
	\caption{Proper Motion for X-Ray Knots of M87}
	\label{table:motion}
	\begin{tightcenter}
	\begin{tabular}{ c c c c c c c }
		\hline
		\hline
		Knot & \multicolumn{2}{c}{Distance$^{a}$} &  $\mu_{\parallel}^{b}$ & $\mu_{\bot}^{c}$ & $\beta_{{\rm app}, \parallel}^{b}$ & $\beta _{{\rm app}, \bot}^{c}$ \\
		& (\arcsec) & (kpc) & (mas yr$^{-1}$) & (mas yr$^{-1}$) & (c) & (c) \\  
		\hline
		HST-1 & 0.99 & 0.080 & $24.1 \pm 1.6$ & $10.9 \pm 0.6$ & $6.3 \pm 0.4$ & $2.9 \pm 0.2$ \\
		D & 2.79 & 0.226 & $9.2 \pm 2.3$ & $0.4 \pm 1.2$ & $2.4\pm 0.6$ & $0.1 \pm 0.3$ \\
		E & 6.14 & 0.497 & $< 5.4$ & $< 3.4$ & $<1.4$ & $<0.8$ \\
		F & 8.41 & 0.681 & $< 5.9$ & $< 3.3$ & $<1.6$ & $<0.9$ \\
		I & 10.98 & 0.889 & $< 7.4$ & $< 4.2$ & $<1.9$ & $<1.1$ \\
		A & 12.40 & 1.004 & $< 2.5$ & $< 1.4$ & $<0.7$ & $<0.4$ \\
		B & 14.48 & 1.173 & $< 5.1$ & $< 3.0$ & $<1.4$ & $<0.8$ \\
		C & 18.20 & 1.474 & $< 9.5$ & $< 5.4$ & $<2.5$ & $<1.4$ \\
		\hline
	\end{tabular}
	\end{tightcenter}
	{${}^{a}$Distance from the AGN as measured from centroid position using 2017 epoch\\
	${}^{b}$Measured parallel to the jet axis\\
	${}^{c}$Measured perpendicular to the jet axis\\
}
\end{table*}

Having found evidence of movement in the difference map, several approaches were tried for determining the proper motions of the knots. First, akin to the method used to align the images in Section~\ref{sect:observation}, the cross-correlation of knot images from the two epochs was generated to determine relative offsets. A unique region was defined for each knot where all other jet features and background point sources were masked. Each region was varied by size, position, and orientation multiple times to ensure the results were not biased by the initial region selection. A two-dimensional cross-correlation function was calculated between epochs for each knot, where all possible alignments of the two images were sampled in the cross-correlation function. The maximum peak of the function was fitted with a two-dimensional Lorentzian profile to determine the relative image offsets. 

Using the cross-correlation method for each knot, a clear offset was measured for Knot D with a radial shift $\Delta_{\parallel}$ = $0.046 \pm 0.011 \arcsec$ and transverse shift $\Delta_{\bot}$ = $0.002 \pm 0.006 \arcsec$. This shift equals a speed of $\Delta_{\parallel} = 9.2 \pm 2.3 \rm\ mas\ yr^{-1}$ and $\Delta_{\bot} = 0.4 \pm 1.2\rm\ mas\ yr^{-1}$ , or $\beta_{\parallel} = 2.4 \pm 0.6$ and $\beta_{\bot} = 0.1 \pm 0.3$. Unfortunately, the cross-correlation method failed to identify proper motions for the remaining knots in the jet due to low count statistics or issues separating the knot's emission from adjacent sources, the latter of which is discussed further in Section~\ref{sect:motionimage}. 

Having attempted measurement of the proper motions within the jet, upper limits were calculated for the  remaining knots using the standard deviation in centroid positions. This technique was chosen because it provides accurate uncertainties for knots with low count statistics. Additionally, the centroid uncertainties were consistent with the cross-correlation uncertainties of knots with high count rates, such as Knot A. To measure the centroid uncertainty, a region was again defined surrounding each knot based on the criteria described in Section~\ref{sect:observation}, and the \ciao{} task {\tt dmstat} was used to locate the centroid position. The centroid position, $\bar{x}$, of a region is defined as the count-weighted mean,
\begin{equation}
\bar{x} = \sum_{{\rm pixel}\ i} \frac{n_{i} x_{i}}{N}, 
\label{eq:average}
\end{equation}  
where $x_i$ is the position of pixel $i$, $n_i$ is the number of
events in this pixel, and $N$ is the total event counts for the region. The standard deviation in the centroid position is  estimated as 
\begin{equation}
\Delta x = \sqrt{ \frac{1}{N(N-1)} \left(\sum_{{\rm pixel}\ i} n_{i}x_{i}^2 - N\bar{x}^2\right)} .
\label{eq:uncertainty}
\end{equation} 
For each knot, the statistical uncertainties of both epochs were combined in quadrature to estimate the total proper motion uncertainty. See Table~\ref{table:motion} for the proper motion upper limits for the remaining knots in the X-ray jet. 

\subsection{Motion Measurements from Comparisons with Synthetic Observations}
\label{sect:motionimage}

 \begin{figure*}
	\begin{tightcenter}
	\includegraphics[width=0.95\textwidth]{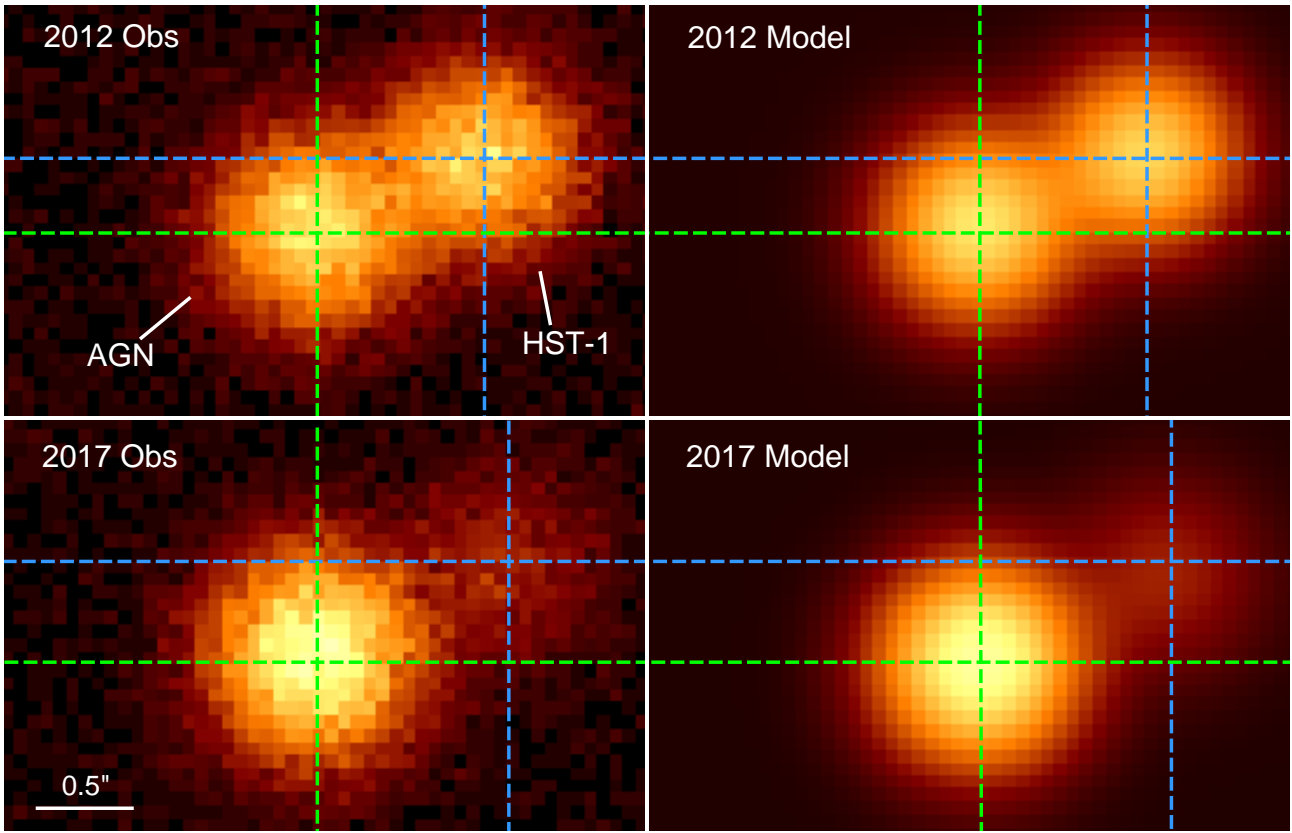}
	\end{tightcenter} 		
      \caption{Comparison of the observed AGN/HST-1 region of 
      		the M87 jet (left column) to simulated images generated from the 
		method discussed in Section~\ref{sect:motionimage} (right column). 
		The images are binned on a scale of $0.066\arcsec \rm\ pix^{-1}$
		and are in units of counts. The dashed lines illustrate the center points of
		the AGN (green) and HST-1 (blue) from the different model fits. 
		Comparison of the 2012 and 2017 epochs shows the proper motion of 
		HST-1 with respect to the AGN.}
	\label{fig:modelfit}
\end{figure*}

Using the methods discussed in Section~\ref{sect:motioncc}, direct measurement of proper motion was not obtained for HST-1. Due to HST-1's close proximity to the AGN and its decrease in brightness of 73\% from 2012 to 2017, we were unable to establish good determinations of its centroid locations for the two epochs. A different measurement technique was required that would be insensitive to significant brightness changes at small separation angles. We therefore opted to generate a model image of the AGN/HST-1 complex from which we may measure offsets between the two epochs.

To begin, a 4\arcsec{} $\times$ 4\arcsec{} region surrounding the AGN/HST-1 complex was extracted from the image for each epoch, avoiding background point sources and other jet features. Both observations were binned at a subpixel size of 0.066\arcsec{} pix$^{-1}$. Synthetic images of the region were generated by modeling the AGN and HST-1 each as a two-dimensional Gaussian where the $x$-position, $y$-position, amplitude, and FWHM parameters were allowed to freely vary. A constant background component was also included in the model. The model was fit to each epoch using the \sherpa{} fitting package \citep{Freeman2001, Fruscione2006}, with the Nelder--Mead method using Cash statistics. The observations are compared to the best-fit synthetic images in Figure~\ref{fig:modelfit}, while the final fitted parameters for the model are available in Appendix~\ref{appen:model}. For both epochs, the model image agrees well with the observed structure.

From the modeled results, centroid coordinates for the AGN and HST-1 were obtained to high accuracy. As any motion of the AGN is assumed to be unresolved \citep{Kovalev2007}, the AGN was treated as a stationary point in the system. We may therefore measure the distance from the AGN to HST-1 for both epochs and compare the results to determine the motion of the knot. In comparing the relative distances, HST-1 was found to move $\Delta_{\parallel}$ = $0.121 \pm 0.008 \arcsec$ and $\Delta_{\bot}$ = $0.054 \pm 0.003 \arcsec$ with respect to the jet axis over the 5 yr time span. This shift corresponds to a proper motion speed of $\Delta_{\parallel}$ = $24.1 \pm 1.6 \rm\ mas\ yr^{-1}$ and $\Delta_{\bot}$ = $10.9 \pm 0.6\rm\ mas\ yr^{-1}$, or $\beta_{\parallel} = 6.3 \pm 0.4$ and $\beta_{\bot} = 2.9 \pm 0.2$.

\section{Discussion}
\label{sect:discuss}

\subsection{Comparison of Proper Motion Measurements to {\it HST} Observations}
\label{sect:compare}

 \begin{figure*}
	\begin{tightcenter}
	\includegraphics[width=0.99\textwidth]{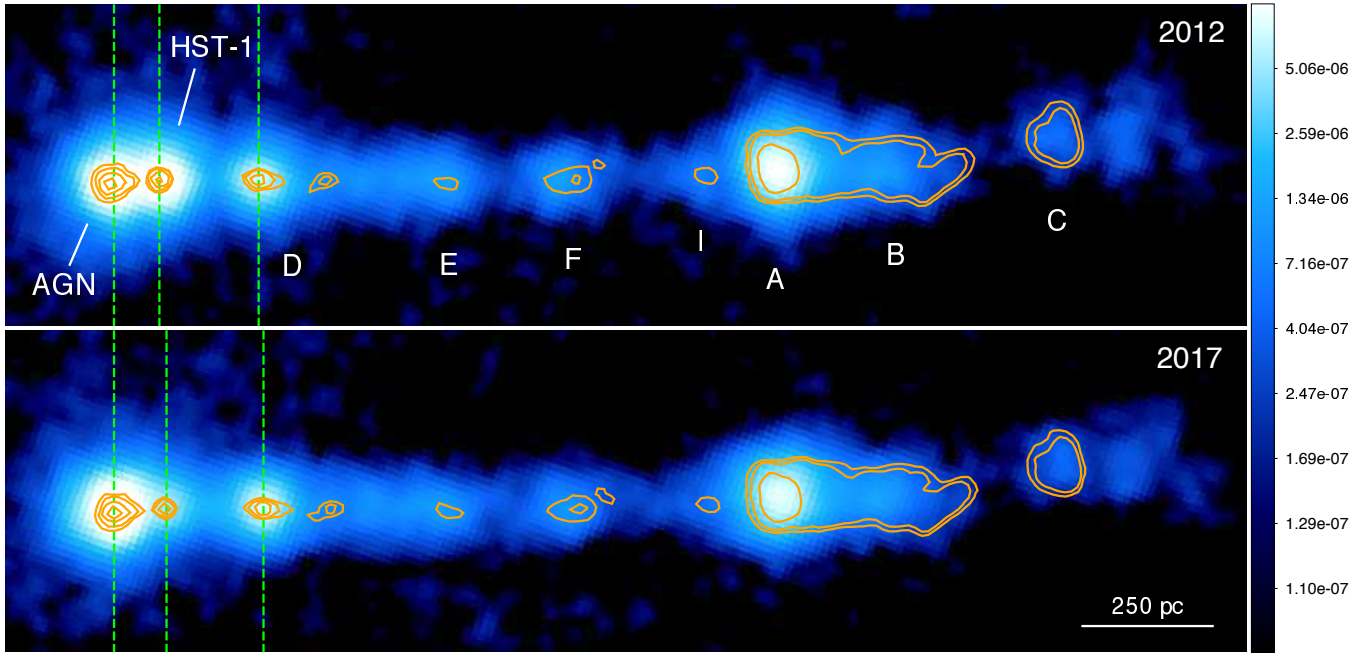}
	\end{tightcenter} 		
      \caption{0.08--10.0 keV HRC-I \chandra{} images of M87 from 2012 and 2017 overlaid
      		with \hst{} WFC3/UVIS F275W contours observed during the same years.
		The X-ray and optical emissions are spatially consistent in both epochs, 
		indicating both emission bands are generated from the same set of sources. 
		Also overlaid are vertical lines (green, dashed) which correspond to the X-ray 
		centroid position of the AGN, HST-1, and Knot D in each epoch. 
		A shift is visible between the epochs, and the shift agrees with the observed 
		optical band proper motion.}
	\label{fig:chandra_hst}
\end{figure*}

The proper motion results from Section~\ref{sect:motion} provide two equally probable interpretations of the system. The first interpretation is that the measurements are due to motion of the X-ray knots, while the second assumes that brightening and/or fading of substructure within the knots gives the appearance of motion at \chandra's resolution. Validation that the observed shifts are indeed motion requires the \chandra{} observations to be compared with contemporaneous higher-resolution observations to ensure that: (1) the X-ray data traces the emission regions to the desired spatial accuracy, and (2) the measured proper motion is consistent between the high- and low-resolution datasets. 

To test whether the knot motions reflect material motion in the jet, the \chandra{} observations were compared to contemporaneous, archival \hst{} images of M87. Previous analyses of M87 with \hst{} have shown evidence of proper motions for all knots within the jet \citep{Biretta1995, Biretta1999, Meyer2013}, and the spatial resolution of \hst{} is a factor of $\sim 3$ better than HRC-I, making it an excellent reference for comparison. Archival WFC3/UVIS \hst{} images at F275W were used (PropIDs \#12989 and \#14618), where the pivot wavelength for the filter is 270.4\,nm. These \hst{} observations, obtained on 2012 December 25 and 2017 March 3, were selected to be the closest match to the \chandra{} epochs (2012 April 14 and 2017 March 02) Figure~\ref{fig:chandra_hst} compares the \chandra{} observations with the \hst{} data included as overlays. Visual comparison of the UV and X-ray knots shows their positions to be consistent at the two epochs. Centroid positions of the knots were compared between the X-ray and UV datasets, with both in agreement to the positions reported in Table~\ref{table:motion} to within the centroid position error determined from the HRC-I observations. Figure~\ref{fig:chandra_hst} also illustrates the centroid position of the X-ray knots for which proper motion was clearly detected, HST-1 and Knot D, overlaid as dashed lines. Using the lines as reference, a shift in the centroid position is observed for each knot, and these position shifts are visually consistent with the proper motions from \hst. 

Beyond visual comparisons, we may compare the measured X-ray proper motions to previous results from \hst{} observations. M87 proper motion measurement from \cite{Biretta1999} and \cite{Meyer2013} were selected for comparison, as shown in Figure~\ref{fig:proper_motion}. \hst{} and \chandra{} proper motion measurements both parallel and perpendicular to the jet are broadly consistent for the knots examined. The only notable discrepancy observed is for the transverse motion of HST-1, which is discussed further in Section~\ref{sect:overview}. Altogether, these results indicate that the X-ray and optical/UV knots track the same physical locations in the jet of M87, to within the spatial accuracy of the X-ray measurements.

\subsection{Overview of Proper Motions}
\label{sect:overview}

 \begin{figure}
	\begin{tightcenter}
	\includegraphics[width=0.46\textwidth]{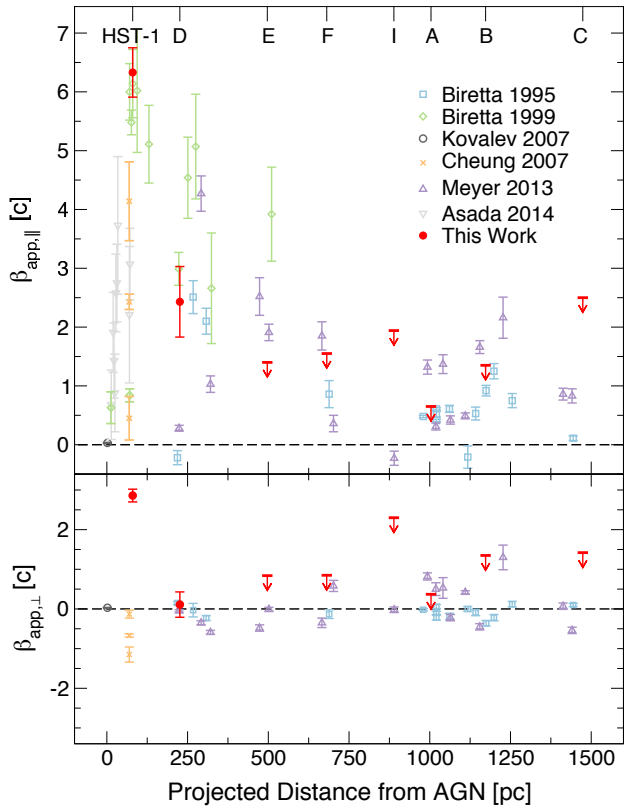}
	\end{tightcenter} 		
      \caption{Apparent speeds in M87 parallel to the jet (upper panel) and perpendicular to the jet (bottom panel) versus projected distance from the AGN in parsecs. The new \chandra-based limits from this paper are shown in red. Previous measurements are taken from \citet{Biretta1995,Biretta1999,Kovalev2007,Cheung2007,Meyer2013,Asada2014}.}
	\label{fig:proper_motion}
\end{figure}

The jet velocity of M87 has been extensively studied on both parsec and kiloparsec scales using radio and optical observations. These measurements include thorough proper motion analyses of the knots within the jet. It is therefore interesting to compare the proper motion results from X-rays with independent measurements from other wavelength bands. See Figure~\ref{fig:proper_motion} for a comparison of reported proper motions in M87, including the X-ray results reported here.  

Among the knots investigated, HST-1 is by far the fastest-moving in all bands observed. X-ray and optical observations both show HST-1 to move at a speed of $\sim$$6c$ along the jet axis \citep{Biretta1999}, while VLBI observations place an upper limit of $5c$ \citep{Cheung2007}. Given the reasonable agreement in speeds of HST-1 before and after its 2002-2008 flaring period \citep{Harris2003, Harris2009}, the flaring event appears to have had little impact on the average proper motion of the knot along the jet axis. 

Examination of the measured transverse proper motions for HST-1 shows a speed of $+2.9 \pm 0.2c$ for X-rays, while the prior VLBI observations show an average transverse speed of $-0.5c$. Transverse motion measurements were not reported for the \hst{} data. The transverse motion detected from X-rays may be due to variations in brightness of substructure not resolvable with \chandra, though it is surprising that the VLBI, which is able to resolve significantly more internal structure of the knot, indicates motion in the opposite direction. In contrast, the centroid positions of HST-1 agree well between \chandra{} and \hst{} at both epochs (Figure~\ref{fig:chandra_hst}), and there is no evidence of brightness changes in the optical/UV or of substructure within it that would be unresolved by \chandra. This comparison supports the conclusion that optical/UV and X-rays track the same physical feature in the jet. Follow-up observations with high-resolution interferometry, such as the Atacama Large Millimeter/submillimeter Array (ALMA) or Square Kilometer Array (SKA), would help determine whether the measured transverse shift is proper motion or is instead related to substructure changes at spatial scales smaller than the resolution of \hst. Additional \chandra{} observations will be useful to further verify the transverse motion.

Moving down the jet, Knot D is demonstrably slower than HST-1 as its proper motion parallel to the jet is only $2.4 \pm 0.6c$ in X-rays. This result is in excellent agreement with the average motion from VLA and \hst{} observations \citep{Biretta1995, Biretta1999, Meyer2013}. Transverse motions also agree well, showing no evidence for motion away from the jet axis. Knot D has shown remarkable consistency in velocity over the 23 yr probed through proper motion observations. 

Upper limits on proper motion for the remaining knots were also extracted from the X-ray data and may be compared to previous estimates. The X-ray result for Knot E places limits 50-75\% lower than those found from \hst{} data in \cite{Biretta1999} and \cite{Meyer2013}.  This is unsurprising given that the \hst{} analyses probed fast-moving edges of the knot, whereas the X-ray data provide the average motion due to the lower spatial resolution of \chandra. Limits established for Knots F, A, and B are each consistent with the average proper motions reported from \hst{} and VLA observations. The less restrictive limits for Knots I and C also agree with independent measurements, as expected given the poor X-ray count statistics, and subsequent large centroid uncertainty, for each feature. Follow-up \chandra{} observations will provide larger time differences between epochs that will reduce the X-ray limits and, ultimately, resolve proper motions for the remaining knots. Based on average speeds from other wavelengths, \chandra{} observations as early as 2020 would provide a sufficient baseline to detect proper motion in Knots E, F, A, and B.

\subsection{Adiabatic Cooling}
\label{sect:adiabatic}

The decreases in brightness observed for HST-1 and Knot A (Section~\ref{sect:diffmap}) might be explained by adiabatic expansion of the radiating region over the examined time interval. As described in \cite{Snios2019}, synchrotron flux $F_E$ at a fixed energy will scale as $F_E \propto V^{-2p/3}$ under isotropic expansion, where $V$ is the volume of the emitting region and the electron energy distribution has the form $dN/d\gamma \propto \gamma^{-p}$ for electron Lorentz factor $\gamma$. 

If the X-ray fading is due to adiabatic losses, we should expect to see a similar reduction in brightness in other bands.  The 2012 and 2017 \hst{} observations from Section~\ref{sect:compare} were therefore studied for brightness changes comparable to those observed in X-rays. Using the archival \hst{} observations, HST-1 and Knot A were each found to have the same UV intensity in both epochs to within their respective uncertainties. The lack of evidence for any change in the optical/UV brightness of the knots strongly disfavors adiabatic loss as the primary cause of the fading observed in both HST-1 and Knot A.

\subsection{Synchrotron Cooling}
\label{sect:synchrotron}

Beyond adiabatic expansion, synchrotron cooling is another potential mechanism that may explain the observed decreases in brightness of the knots. The synchrotron cooling rate for an X-ray knot may be estimated with the Kardashev--Pacholczyk (KP) model \citep{Kardashev1962, Pacholczyk1970}. As discussed in \cite{Snios2019}, the cooling rate is maximized when the particles move perpendicular to the magnetic field and particle scattering is assumed to be negligible. Under these assumptions, we can estimate the minimum magnetic field strength required to account for the observed fading of HST-1 and Knot A. The initial electron distribution for each knot was taken to have the form $dN/d\gamma = K \gamma^{-p}$, with  $p=3.6$ for HST-1 and $p=4.2$ for Knot A based on independent measurements \citep{Kataoka2005, Perlman2005}. A maximum Lorentz factor of $10^9$ was used for the model, making the initial X-ray spectrum a power law to well above the energies accessible to \chandra. The electron distribution was evolved, allowing for the effects of synchrotron cooling, over the 5 yr time span of the X-ray observations to determine the final X-ray spectrum, which was folded with the HRC-I response to determine the change in count rate over this interval. The magnetic field strength was adjusted to obtain the observed reduction in count rate. Relativistic effects (time dilation and light travel delay), which are discussed further below, were ignored initially for simplicity. Under these assumptions, HST-1 required a minimum magnetic field strength of $\sim$\,800~$\mu$G, while Knot A required a minimum of $\sim$\,250~$\mu$G. These values are consistent with estimates of the equipartition magnetic field \citep{Harris2003, Kataoka2005, Stawarz2005, Harris2006, Harris2009}. Under synchrotron cooling, the optical/UV emission of the knots should remain constant over the 5 yr span of these observations, consistent with our findings for the archival HST observations (Section~\ref{sect:adiabatic}). 

As noted in Section~\ref{sect:adiabatic}, the constant optical/UV knot emission argues against adiabatic loss models. Our determination of the minimum magnetic field strength from the synchrotron cooling model relies on the same electron population being responsible for the X-ray emission at both observing epochs, requiring the knot material to move at relativistic speeds between the observations. This is noteworthy because it implies that the speeds of the jet knots reflect bulk relativistic motion of jet plasma, not just of a disturbance, such as a wave or shock front in the jet. To avoid this conclusion, the jet plasma emitting at the initial knot position would have to cool even faster than assumed, requiring a substantially greater magnetic field. Since the required field strength would then be larger than the equipartition value, this seems unlikely. The simplest conclusion is that the motion of the jet knots directly reflects the bulk speed of the jet plasma.

For a knot moving at speed $\beta c$, the proper duration of a process, $t_p$, is boosted by time dilation to $\gamma t_p$, where $\gamma = 1 / \sqrt{1 - \beta^2}$, in the rest frame. Light travel delay reduces the duration measured by a stationary observer by an
additional factor of $1 - \beta \cos\theta$, where $\theta$ is the angle between the velocity of the knot and the observer's line of sight ($\theta = 0$ for a knot approaching directly). Thus, the stationary observer measures the duration
\begin{equation}
t_{o} = \gamma(1-\beta cos \theta)t_{p} = t_{p}/\delta\, 
\label{eq:time_dilation}
\end{equation} 
where $\delta$ is the jet Doppler factor. The synchrotron cooling time of electrons in the knot that radiate photons of energy $\epsilon_p$ scales as $t_p \propto (\epsilon_p B^3)^{-1/2}$ and the energy of those photons in the observer's frame would be $\epsilon_o = \epsilon_p \delta$. Combining these scalings, the relativistic correction to our estimate of the minimum magnetic field strength in the knot's rest frame would be a factor of $\delta^{-1/3}$. While the Doppler factor is poorly constrained, the apparent knot speed, $\beta_{p} = \beta \sin\theta / (1 - \beta \cos\theta)$ is known, placing a joint constraint on $\beta$ and $\theta$. For fixed $\beta$, the expression for $\beta_p$ is maximized as a function of $\theta$ when $\cos\theta = \beta$, therefore the minimum possible value for the knot momentum is $\beta\gamma = \beta_{p}$. Although the Doppler factor could still have any positive value, if higher knot momenta are less likely, reasonable values of the Doppler factor will be comparable to the value corresponding to the minimum momentum, i.e., $\delta_m = \sqrt{1 + \beta_{p}^2}$.

Applying the Doppler factor $\delta_m$ to the observed 5 yr time interval increases the cooling time for HST-1 to 35.1 yr and Knot A to 6.2 yr. Using the revised timespans, HST-1 required a minimum magnetic field strength of $\sim$\,420~$\mu$G, and Knot A required a minimum of $\sim$\,230~$\mu$G. These revised lower limits are again consistent with previous equipartition magnetic field estimates, and the synchrotron cooling agrees with the constant optical/UV intensity observed via \hst. Given the consistency between observations and the models in both cases tested, synchrotron cooling is therefore the most probable mechanism to explain the observed brightness decreases for HST-1 and Knot A. 

\section{Conclusions}
\label{sect:conclusions}

\chandra{} HRC observations were analyzed for evidence of proper motions and brightness variations within the X-ray jet of the radio galaxy M87. Observations from 2012 and 2017 were coaligned to high accuracy, and a difference map was generated between the epochs. Visual evidence of proper motions was seen in the difference map over the 5 yr time span, so follow-up measurements were performed using three different techniques, as appropriate, to determine the proper motions of the X-ray knots. From these methods, proper motions were directly measured for the innermost knots, HST-1 and Knot D. HST-1 was measured to have a projected speed of $24.1 \pm 1.6 \rm\ mas\ yr^{-1}$, or $6.3 \pm 0.4 c$, parallel to the jet and $10.9 \pm 0.6  \rm\ mas\ yr^{-1}$,  or $2.9 \pm 0.2c$, perpendicular to the jet. Knot D was measured at $9.2 \pm 2.3 \rm\ mas\ yr^{-1}$, or $2.4\pm 0.6c$, parallel and $0.4 \pm 1.2  \rm\ mas\ yr^{-1}$,  or $0.1 \pm 0.3c$, perpendicular to the jet. Upper limits on proper motions for the remaining knots in the jet were established using centroid analysis. 

Results from this work were compared with previous measurements from other energy bands. Archival \hst{} observations were overlaid with the \chandra{} observations, and excellent agreement was observed in spatial positions for both epochs. The superb alignment indicates that the X-ray and optical/UV knots track the same physical locations in the jet of M87. Proper motion measurements from the X-ray data were compared with results from UV, optical, and radio observations. The X-ray-based proper motions were consistent with independent measurements for all knots within the jet. Based on the average speeds from other wavelengths, follow-up \chandra{} observations as early as 2020 would be sufficient detect proper motions for the majority of remaining knots in the system (Knots E, F, A, and B).

Brightness variations up to 73\% were observed for the knots in M87, with HST-1 demonstrating the most significant change over the 5 yr time span analyzed. Potential emission mechanisms that may explain the observed fading of knots were tested. Archival \hst{} data were compared against the X-ray data, and the knots in M87 remained constant in brightness at optical/UV wavelengths. The lack of optical/UV brightness changes, together with the significant X-ray changes, disfavor adiabatic loss as the primary cause of fading. Synchrotron cooling was also studied as a potential fading mechanism. Lower limits on magnetic field strengths were estimated from the synchrotron models, providing minimum magnetic field strengths of $\sim$\,420~$\mu$G for HST-1  and $\sim$\,230~$\mu$G for Knot A. These limits are consistent with equipartition magnetic field estimates, and the synchrotron cooling model also agrees with the constant optical/UV intensity observed via \hst. Synchrotron radiation is therefore the most likely cause of fading. The preference for the synchrotron cooling model together with the agreement in positions and speeds between the X-ray and optical/UV emission provide a strong case that the observed knot speeds reflect the relativistic speed of the jet plasma, not just of a disturbance propagating along the jet.
\\
\acknowledgements{
Support for this work was provided by the National Aeronautics and Space Administration through \chandra{} Award Number G07-18104X issued by the {\it Chandra X-ray Observatory} Center, which is operated by the Smithsonian Astrophysical Observatory for and on behalf of the National Aeronautics Space Administration under contract NAS8-03060. P.E.J.N. and R.P.K. were supported in part by NASA contract NAS8-03060.
}

\software{
\ciao{} v4.10 \citep{Fruscione2006},
\sherpa{} v1 \citep{Freeman2001}
}

\bibliographystyle{aasjournal}
\bibliography{all_data}

\appendix 

\section{Model Image Parameters} 
\label{appen:model}

This appendix features the parameters used for the model image fits described in Section~\ref{sect:motionimage}. The model is a combination of two, two-dimensional Gaussians together with a constant background emission. Best-fit parameters are provided in Table~\ref{table:modelparameters}. The table includes the epoch-fitted FWHM for the AGN FWHM$_{\rm AGN}$, $x$-position of the AGN $x_{\rm AGN}$, $y$-position of the AGN $y_{\rm AGN}$, amplitude of the AGN $A_{\rm AGN}$, FWHM for HST-1 FWHM$_{\rm HST\mbox{-}1}$, $x$-position of HST-1  $x_{\rm HST\mbox{-}1}$, $y$-position of HST-1 $y_{\rm HST\mbox{-}1}$, amplitude of HST-1 $A_{\rm HST\mbox{-}1}$, and the background amplitude $A_{\rm bkgd}$. The directions of $(x,y)$ correspond to the directions of R.A. and decl., respectively. Positions and distances are in units of pixels, where $1 {\rm\ pixel} = 0.066\arcsec$. Amplitudes are in units of counts. We note that the differences in AGN coordinates between epochs are due to different coordinate map definitions for the two fits and are not representative of astrometric shifts.

\begin{table}[h]
	\caption{Model Fit Parameters for the AGN/HST-1 Region}
	\label{table:modelparameters}
	\begin{tightcenter}
		{\footnotesize
		\begin{tabular}{ c c c c c c c c c c}
		\hline
		\hline
		& FWHM$_{\rm AGN}$ & $x_{\rm AGN}$ & $y_{\rm AGN}$ 
			& $A_{\rm AGN}$ & FWHM$_{\rm HST\mbox{-}1}$ &
			$x_{\rm HST\mbox{-}1}$ & $y_{\rm HST\mbox{-}1}$ & 
			$A_{\rm HST\mbox{-}1}$ & $A_{\rm bkgd}$ \\
		Epoch & (pix) & (pix) & (pix) & (cts) & (pix) & (pix) & (pix) & (cts) & (cts) \\
		\hline
		2012 & $10.70\pm0.05$ & $465.08\pm0.06$ & $187.01\pm0.05$ & 
			$100.2\pm1.9$ & $10.45\pm0.05$ & $478.04\pm0.06$ & $192.52\pm0.05$
			& $88.7\pm1.1$ & $0.98\pm0.02$\\
		2017 & $11.07\pm0.05$ & $465.74\pm0.04$ & $187.21\pm0.04$ & 
			$132.4\pm1.4$ & $11.07\pm0.05$ & $480.17\pm0.14$ & $194.09\pm0.12$
			& $18.7\pm0.5$ & $0.86\pm0.02$\\	
		\hline
	\end{tabular}
	}
	\end{tightcenter}
\end{table}

\end{document}